\documentclass[%
 reprint,
 amsmath,amssymb,
 aps,
]{revtex4-2}

\usepackage{graphicx}
\usepackage{dcolumn}
\usepackage{bm}
\usepackage{capt-of}

\begin{document}
\raggedbottom

\preprint{APS/123-QED}

\title{Localization Transitions in a Half-Filled Helical Aubry-Andr\'e Model}

\author{Taylan Yildiz}
 \affiliation{%
Department of Physics, Bilkent University, 06800 Ankara, T{\"u}rkiye}

 \author{Balázs Hetényi}

\affiliation{Department of Theoretical Physics, Budapest University of Technology and Economics, H-1111 Budapest, Hungary\\}
\affiliation{MTA-BME Lend\"{u}let “Momentum” Open Quantum Systems Research Group, Institute of Physics, Budapest University of Technology and Economics, M\H{u}egyetem rkp. 3, H-1111 Budapest, Hungary\\}
\affiliation{Institute for Solid State Physics and Optics, HUN-REN Wigner Research Centre for Physics,  H-1525 Budapest, P. O. Box 49, Hungary\\}

\author{B. Tanatar}%
 \email{tanatar@fen.bilkent.edu.tr}
 
\affiliation{%
Department of Physics, Bilkent University, 06800 Ankara, T{\"u}rkiye}

\date{\today}

\begin{abstract}

We study localization in a one-dimensional quasiperiodic lattice obtained by extending the 
Aubry-Andr\'e model with an additional $N$th-neighbor hopping term of strength $J_{N}$. This long-range tunneling couples successive windings of an effective helical chain and introduces a second control parameter beyond the quasiperiodic potential strength $\Delta$. Working with noninteracting fermions (typically at half filling), we diagnose the delocalization-localization transition using extensions of the modern theory of polarization. Specifically, we compute the polarization amplitudes of the many-body Slater-determinant ground state and construct a geometric Binder cumulant from polarization amplitudes. The critical potential where the localization transition happens is extracted from the sign change (zero crossing) of the geometric Binder cumulant. We map critical potential as a function of $J_N$ and the helical range $N$, finding that stronger helical hopping generally stabilizes the extended phase (shifting critical potential upward), while the $N$-dependence can display pronounced commensurability-induced spikes. We further compare the geometric Binder cumulant with the Fermi gap, which remains near zero at small values of potential and opens in the same parameter regime where the geometric Binder cumulant departs from extended phase. Finally, to take a controlled thermodynamic limit along Fibonacci system sizes, we employ a Zeckendorf-shift construction that fixes the many-body sector consistently as system size goes to infinity.
\end{abstract}

\maketitle

\section{Introduction}

The delocalization-localization transition (DLT) in condensed matter physics has been a topic of intense interest for many decades. Anderson localization (AL) is a localization phenomenon that happens in the presence of random onsite disorder \cite{1}. In such systems, a DLT is only found in three dimensions, in one, or two dimensions extended states are absent \cite{2,3}. Quasiperiodic systems however can exhibit extended states even in one dimension. The best known example of a quasiperiodic lattice is the 
Aubry-Andr{\'e} (AA) model where an incommensurate onsite modulation drives a self-dual DLT at a critical potential strength \cite{4,5,6}. Quasiperodic models, especially the AA model,  offer a controlled setting to study the DLT, mobility edges and finite-size effects \cite{7,8,9,10,11,12,13,14,15}. Extensions of the AA model are also extensively studied in condensed matter physics including those with longer-range tunneling \cite{16,17,18,41}, many-body extensions \cite{19,20,21}, multiple transitions \cite{,22,23,24,25,26,27,28,29} and non-Hermitian generalizations \cite{30,31,32,33}. Because the AA model and its extensions are made with incommensurate lattice potentials, they are experimentally feasible and directly relevant to engineered lattices in cold atoms, optical lattices and electrical circuits \cite{34,35,36,37,38,39,40}. 

In this work we study localization in a one-dimensional quasiperiodic lattice obtained by extending the AA model with an additional \(N\)th-neighbor hopping process of strength \(J_N\).
This term couples sites separated by a fixed range \(N\) and can be viewed as tunneling between successive windings of an effective helical chain \cite{42,43,44,45}. When the lattice is wrapped so that each winding contains \(N\) sites, the long-range hopping connects neighboring windings while the original nearest-neighbor hopping propagates along the winding.

The resulting helical AA model introduces a second tuning parameter \(J_N\) beyond the quasiperiodic potential strength, enabling a systematic exploration of how long-range tunneling reshapes the localization transition.
A key motivation is that adding a structured long-range coupling can qualitatively modify transport and spectral properties, and can also generate pronounced commensurability effects when the helical range \(N\) becomes arithmetically compatible with the system size.

A practical challenge is that many localization diagnostics are formulated in real space and implicitly assume open boundaries, while quasiperiodic models are often studied with periodic boundary conditions to minimize edge effects and to employ rational approximants of the incommensurate modulation.
Under periodic boundary conditions, the position operator is not a well-defined observable on a ring \cite{46}, so directly using moments of position as localization indicators is subtle.
We overcome this difficulty by employing the modern theory of polarization \cite{47,48,49,50,51,52,53}, where the relevant object is a twist (or polarization) operator whose expectation value remains well-defined on a ring and sharply distinguishes localized from extended many-body states.
For a non-interacting fermionic ground state, the polarization amplitudes \(Z_q\) can be evaluated efficiently as determinants of overlap matrices constructed from the occupied single-particle orbitals.
From \(|Z_1|\) and \(|Z_2|\) one obtains approximate second and fourth moments of the centered polarization distribution, which in turn define the geometric Binder cumulant \cite{58,59} \(U_4\).

The geometric Binder cumulant \cite{54,55,56} provides a simple and robust criterion for the delocalization-localization trasnsition:
as the quasiperiodic potential \(\Delta\) is increased, \(U_4(\Delta)\) evolves from an extended-phase and changes sign when the ground state becomes localized.
We therefore extract the critical potential from the zero crossing of \(U_4(\Delta)\), and map \(\Delta_c\) as a function of the long-range hopping amplitude \(J_N\) and the helical range \(N\).
We find that increasing \(J_N\) generally stabilizes the extended phase by pushing \(\Delta_c\) to larger values, while the dependence on \(N\) can show pronounced spikes that correlate with commensurability conditions.
To connect the polarization-based diagnosis to spectral properties, we also compare \(U_4(\Delta)\) with the Fermi gap, which remains near zero at small \(\Delta\) and opens in the same parameter regime where \(U_4\) departs from its extended-phase behavior.

Finally, to take a controlled thermodynamic limit in a quasiperiodic setting, we work along Fibonacci system sizes \(L=F_n\), which are natural rational approximants for incommensurate modulations under periodic boundary conditions.
However, specifying only a target filling ratio does not uniquely determine how the many-body sector is approached as \(n\to\infty\).
We therefore employ a Zeckendorf-shift construction \cite{57}: starting from a reference particle number with a fixed Zeckendorf decomposition \cite{60}, we generate particle numbers for larger Fibonacci sizes by shifting the Fibonacci indices.
This defines a unique sequence of many-body sectors and yields a well-defined thermodynamic-limit filling, enabling consistent finite-size scaling of \(U_4\) and of \(\Delta_c\).

The paper is organized as follows. In Sec.~II, we introduce the helical Aubry-Andr\'e model and define the additional $N$th-neighbor hopping term that gives rise to the effective helical geometry. In Sec.~III, we outline the geometric Binder cumulant approach under periodic boundary conditions, summarize the construction of the polarization amplitudes, and explain how the critical quasiperiodic potential $\Delta_c$ is extracted from the zero crossing of $U_4(\Delta)$ In Sec.~IV, we present the finite-size results, including the dependence of $\Delta_c$ on the helical hopping strength $J_N$ and on the winding $N$, and compare the behavior of $U_4(\Delta)$ with the single-particle Fermi gap and the spectral evolution. In Sec.~V, we discuss the Zeckendorf-shift construction and the thermodynamic-limit extrapolation along Fibonacci system sizes, and use this framework to examine which features of the phase boundary persist in the $L\to\infty$ limit. Finally, in Sec.~VI we summarize our main results and conclusions. Technical details concerning the discrete moment expansion and the formulas for $M_2$ and $M_4$ are given in Appendix~A.

\section{Helical Aubry-Andr\'e Model}
\begin{figure}
    \centering
    \includegraphics[width=\linewidth]{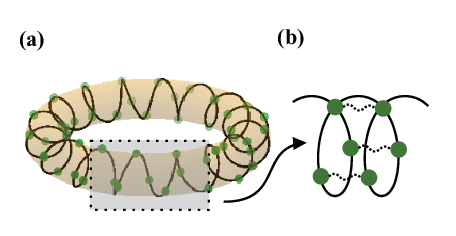}
    \caption{\textbf{(a)} Schematic illustration of a one-dimensional lattice embedded on a toroidal geometry. The sites follow a helical path wrapped around the torus, such that periodic boundary conditions are naturally realized by the closed topology. The shaded region highlights a local segment of the lattice.
\textbf{(b)} Effective representation of the helical structure, where sites on neighboring turns become spatially close. Solid lines denote nearest-neighbor hopping $J$, while dotted links indicate longer-range hopping $J_N$ emerging from the geometric embedding.
}
    \label{fig:fig}
\end{figure}
We consider the Aubry-Andr\'e Hamiltonian given by 
\begin{equation}
    \hat{H}_{\mathrm{AA}}
=-J\sum_{j=1}^{L}\left(c_{j+1}^{\dagger}c_{j}+c_{j}^{\dagger}c_{j+1}\right)
+\Delta\sum_{j=1}^{L}v_j\,n_j,
\end{equation}
where $v_j=\cos(2\pi\beta j)$ is the quasiperiodic part of the on-site potential and $\beta$ is an irrational number which governs the quasiperodicitiy. $\Delta$ and $J$ correspond to potential and hopping strength, respectively. Operators $c_{j}^{\dagger} \;(c_j)$ are the creation (annihilation) operators for fermions.

Now we extend the AA model to a longer range hopping denoted by the $N$th-hopping, 
\begin{equation}
    H_N=H_{\mathrm{AA}}-J_{N}\sum_{j=1}^{L}\left(c_{j+N}^{\dagger}c_{j}+c_{j}^{\dagger}c_{j+N}\right)
\end{equation}
where $J_N$ is the strength of $N$th-neighbor hopping. The $N$th-hopping gives rise to a helical geometry, as illustrated in Fig.\,\ref{fig:fig}. In a helical chain, this long-range hopping is the tunneling between successive helical windings, and with this setting, each winding has $N$ lattice points.

\section{Geometric Binder Cumulant}
For periodic boundary conditions (PBC), the usual position operator is not a well-defined
observable on a ring, hence real-space localization diagnostics (based directly on $x$) are
not immediate. This difficulty is overcome by the modern theory of polarization (MTP),
which replaces $x$ by a polarization operator whose expectation value
remains well-defined under PBC and is sensitive to localization.

We consider a non-interacting fermionic system of size $L$ described by the helical
Aubry-Andr\'e Hamiltonian, introduced in Sec.~II and we implement PBC, i.e.
$c_{L+1}\equiv c_1$ and $c_{i+N}\equiv c_{\text{mod(i+N-1,L)+1}}$ (and similarly for negative indices).
At a given filling $\nu=N_p/L$, the many-body ground state is a Slater determinant of the
$N_p$ lowest-energy single-particle eigenstates,
\begin{equation}
|\Psi_0\rangle = \prod_{\lambda=1}^{N_p} d^\dagger_\lambda |0\rangle,
\qquad
d^\dagger_\lambda=\sum_{j=1}^L \phi_\lambda(j)\,c^\dagger_j,
\end{equation}
where $\phi_\lambda(j)$ are normalized single-particle eigenvectors of the PBC Hamiltonian.

The central object of MTP is the polarization amplitude (twist expectation value)
\begin{equation}
Z_q \equiv \langle \Psi_0| \hat{U}^{(q)} |\Psi_0\rangle,
\end{equation}
where we have 
\begin{equation}
   \hat{U}^{(q)}=\exp\bigg(\frac{2\pi i q}{L}\hat{X}\bigg), \quad q\in\mathbb{Z}
\end{equation}
and $\hat{X}$ is the many-body position operator. For a Slater determinant, $Z_q$ can be evaluated as a determinant of an $N_p\times N_p$
matrix,
\begin{equation}
Z_q = \det\!\big[ U_{\lambda\lambda'}^{(q)} \big], \qquad
U^{(q)}_{\lambda\lambda'} = \sum_{j=1}^{L}
\phi_\lambda^*(j)\, e^{ i \frac{2\pi q}{L} j }\, \phi_{\lambda'}(j).
\label{eq:Zq_det}
\end{equation}
The set $\{Z_q\}$ is the discrete cumulant-generating function of the total
position $X$ of the charge carriers on the ring. The corresponding distribution $P(X)$ may be
obtained with an inverse discrete Fourier transform,
\begin{equation}
P(X)=\frac{1}{L}\sum_{q=0}^{L-1} Z_q \,
\exp\!\left( i \frac{2\pi q}{L} X \right)
\end{equation}

To construct a Binder-type indicator, we first center the distribution by replacing
$Z_q\mapsto |Z_q|$, which removes an overall phase and shifts $P(X)$ to be centered within a
unit cell. From the centered distribution, one defines approximate second and fourth moments
in terms of $|Z_1|$ and $|Z_2|$:
\begin{align}
M_2 &= \frac{L^2}{2\pi^2}\big(1-|Z_1|\big), \label{eq:M2}\\
M_4 &= \frac{L^4}{8\pi^4}\big(|Z_2|-4|Z_1|+3\big). \label{eq:M4}
\end{align}
Finally, the geometric Binder cumulant is defined as
\begin{equation}
U_4 \equiv 1-\frac{1}{3}\frac{M_4}{M_2^2}.
\label{eq:U4_def}
\end{equation}
As the quasiperiodic
potential strength $\Delta$ is tuned, $U_4(\Delta)$ changes sign at the delocalization-localization
transition. Also, we know that for a delocalized ground state which is nondegenerate, $U_4=1/2$ whereas if such a ground state is degenerate $U_4=1/3$. In practice, the critical point $\Delta_c$ is extracted by locating the size-independent crossing (or sign change) of
$U_4(\Delta)$ computed for several system sizes $L$ at fixed filling $\nu$ and fixed helical
parameters $(N,J_N)$.

For a given parameter set $(L,\nu,\Delta,J_N,N)$ the algorithm is:
(i) diagonalize the single-particle PBC Hamiltonian and collect eigenvectors
$\{\phi_\lambda(j)\}$;
(ii) occupy the lowest $N_p=\nu L$ states to form $U^{(q)}$ for $q=1,2$ 
(iii) compute $Z_1,Z_2$, then $M_2,M_4$, and finally $U_4$ using;
(iv) repeat for multiple $L$ and locate $\Delta_c$ from the finite-size behavior of $U_4$.

A technical issue concerns the shape of $P(X)$: for certain parity combinations of particle
number $N_p$ and system size, $L$ the distribution can become bimodal near criticality, in
which case, the above GBC construction is not directly applicable. In computations, we therefore
focus on opposite-parity cases (odd $L$ with even $N_p$, or vice versa), for which $P(X)$
remains unimodal, and the GBC analysis is robust.
\section{Results and Discussion}
We scale energies by $J$ throughout the results and use half-filling whenever the filling ratio is not specified.
\begin{figure}
    \includegraphics[width=\linewidth]{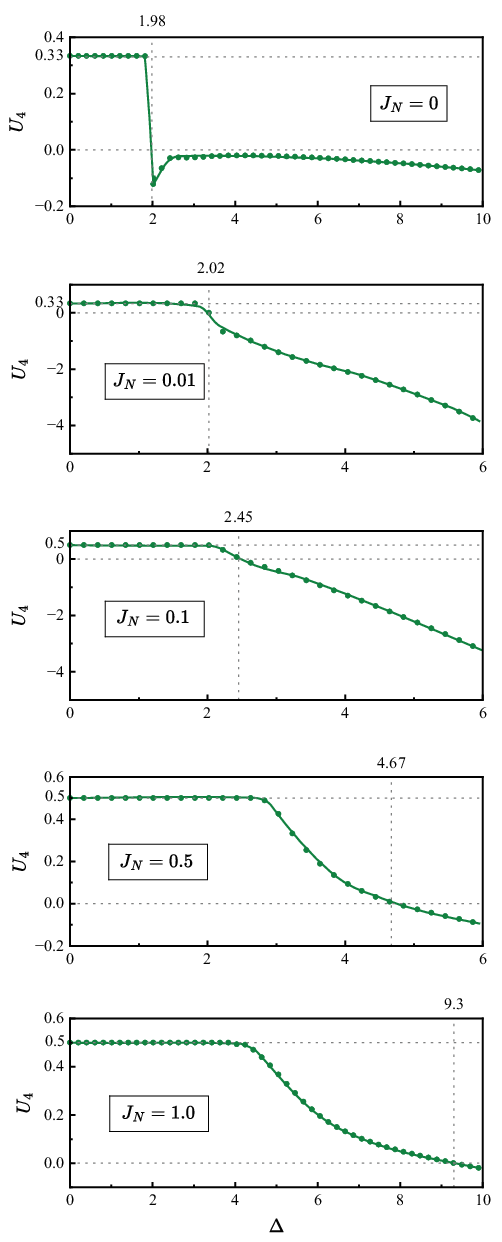}
    \caption{Geometric Binder cumulant $U_4$ as a function of quasiperiodic potential strength $\Delta$ for $L=987$ and $N=40$, shown for several $N$th-neighbor hopping amplitudes $J_N$ (panel labels). Horizontal dotted lines indicate $U_4=1/3,1/2$ and $U_4=0$. The vertical dotted line marks the critical potential $\Delta_c$ defined by the zero crossing of $U_4(\Delta)$.}
    \label{fig:1}
\end{figure}

Fig.\,\ref{fig:1} shows representative traces of the geometric Binder cumulant $U_4(\Delta)$ at fixed helical range $N=40$ and system size $L=987$ for several values of the $N$th-neighbor hopping amplitude $J_N$. For weak quasiperiodic modulation, $U_4$ stays in the extended phase, and we see that $U_4=1/2$ or $U_4=1/3$. If $U_4=1/2$, the ground state is nondegenerate, the distribution of $P(X)$ is flat and if $U_4=1/3$ the ground state is degenerate, the disstribution is a raised cosine. 
As $\Delta$ is increased, $U_4(\Delta)$ decreases and ultimately changes sign, signaling a qualitative change in the polarization fluctuations consistent with the onset of localization.
Following the procedure described in Sec.~III we define the finite-size critical potential $\Delta_c(L)$ as the zero crossing of $U_4(\Delta)$; the corresponding values are marked by the vertical dotted lines and listed above each panel.

In the Aubry-Andr\'e limit $J_N=0$, the sign change occurs at $\Delta_c\simeq 1.98$, consistent with the well-known critical value $\Delta_c=2$ in the units used here.
Turning on a small helical hopping shifts the crossing only mildly ($\Delta_c\simeq 2.02$ for $J_N=0.01$), whereas increasing $J_N$ leads to a pronounced stabilization of the extended phase: the critical potential rises to $\Delta_c\simeq 2.45$ for $J_N=0.1$, to $\Delta_c\simeq 4.67$ for $J_N=0.5$, and reaches $\Delta_c\simeq 9.3$ for $J_N=1.0$.
Physically, the additional long-range tunneling increases the effective kinetic connectivity of the chain (or, in the helical picture, enhances inter-winding transport), so a stronger quasiperiodic modulation is required to suppress coherent motion and drive localization.
Beyond the crossing, $U_4$ becomes negative and decreases further with $\Delta$, indicating increasingly strong localization-induced polarization fluctuations.

These panel-by-panel $U_4(\Delta)$ curves serve as the basic input for constructing the phase boundary.
In the remainder of this section, we repeat the same extraction of $\Delta_c$ over a dense grid of $(J_N,N )$ and for multiple Fibonacci sizes, and then perform finite-size scaling to obtain thermodynamic-limit estimates of $\Delta_c(J _N,N)$ and to identify commensurability-induced anomalies in the $N$ dependence.
\begin{figure}
    \includegraphics[width=\linewidth]{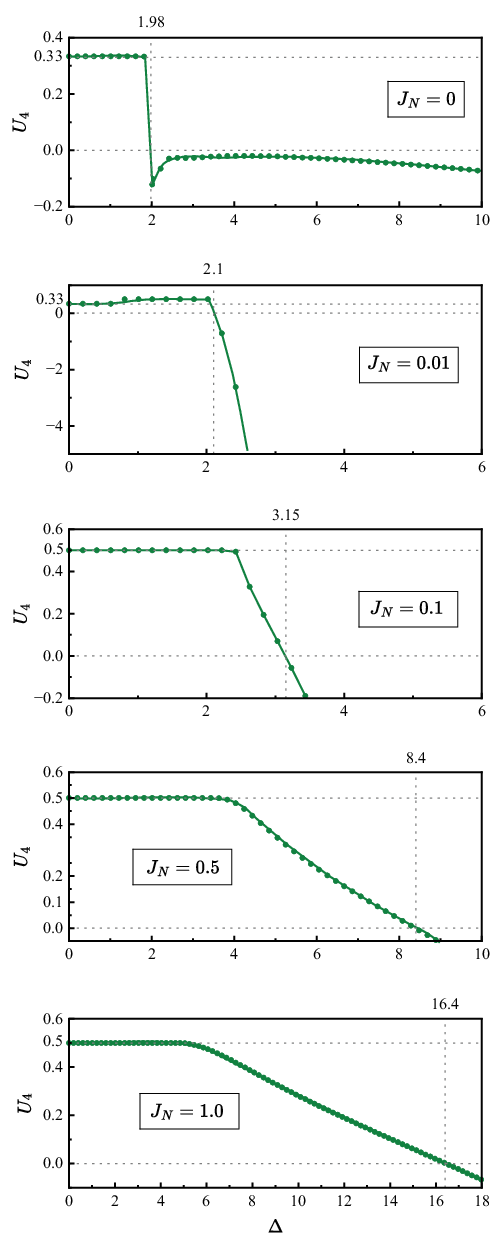}
    \caption{Geometric Binder cumulant $U_4$ as a function of quasiperiodic potential strength $\Delta$ for $L=987$ and $N=100$, shown for several $N$th-neighbor hopping amplitudes $J_N$ (panel labels). Horizontal dotted lines indicate $U_4=1/3,1/2$ and $U_4=0$. The vertical dotted line marks the critical potential $\Delta_c$ defined by the zero crossing of $U_4(\Delta)$.}
    \label{fig:2}
\end{figure}
Fig. \ref{fig:2} demonstrates that this stabilizing effect of helical tunneling persists upon changing the helical range.
Here we keep the system size fixed at $L=987$ but increase the winding to $N=100$.
The qualitative behavior of $U_4(\Delta)$ remains the same across all $J_N$: an extended-phase plateau at small $\Delta$, followed by a monotonic decrease and a sign change at a well-defined $\Delta_c$.
However, the location of the crossing is markedly shifted compared to Fig.\ref{fig:1}
while the AA limit again yields $\Delta_c\simeq 2$ (here $\Delta_c\simeq 1.98$ for $J_N=0$), the critical potential rises much more rapidly with $J_N$ at this larger winding: $\Delta_c\simeq 2.1$ for $J_N=0.01$, $\Delta_c\simeq 3.15$ for $J_N=0.1$, $\Delta_c\simeq 8.4$ for $J_N=0.5$, and $\Delta_c\simeq 16.4$ for $J_N=1.0$.
Thus, at fixed $J_N$ the extended phase is further stabilized as the helical range is increased from $N=40$ to $N=100$.

Within the helical-chain interpretation, this trend is natural.
For larger $N$, the long-range hop bridges a longer spatial separation along the underlying 1D chain, but it still connects adjacent windings of the helix.
As a result, increasing $N$ effectively enhances inter-winding transport relative to the onsite modulation, making it harder for the quasiperiodic potential to pin the ground state.
At the same time, the strong $N$-dependence seen by comparing Figs.\,\ref{fig:1} and \ref{fig:2} highlights that $\Delta_c$ is not determined solely by $J_N$, but by the combined geometry set by  $(N,L)$.

\begin{figure}
    \includegraphics[width=1\linewidth]{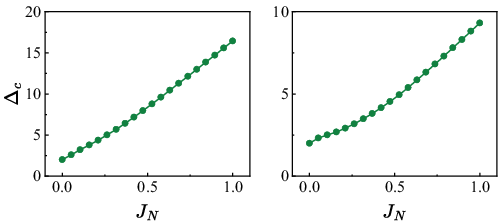}
    \caption{Critical quasiperiodic potential strength $\Delta_c$ as a function of the long-range (helical) hopping amplitude $J_N$ for a chain of length $L=987$ at half filling. The critical value $\Delta_c$ is extracted from the zero crossing of the geometric Binder cumulant $U_4(\Delta)$ upon increasing $\Delta$, with the helical hopping range fixed to (a) $N=100$ (b)$N=40$ and $J_N$ varied linearly from $0$ to $1$.}
    \label{fig:3}
\end{figure}
\subsection{Phase boundary versus helical hopping strength}

Having established how $\Delta_c$ is extracted from the zero crossing of $U_4(\Delta)$, we now summarize the resulting phase boundary in terms of the two helical control parameters.
Figure~\ref{fig:3} shows the critical quasiperiodic potential $\Delta_c$ as a function of the long-range hopping amplitude $J_N$ for a fixed chain length $L=987$ at half filling, for two representative windings: (a) $N=100$ and (b) $N=40$.
In both cases, $\Delta_c(J_N)$ increases monotonically with $J_N$, demonstrating that helical tunneling stabilizes the extended phase: a larger quasiperiodic modulation is required to drive the ground state into the localized regime when the additional $N$th-neighbor channel is present.
The growth is substantially stronger for $N=100$ than for $N=40$, consistent with the comparison between Figs.~\ref{fig:1} and \ref{fig:2} and highlighting that the influence of $J_N$ depends on the helical geometry set by $N$.

\begin{figure}
    \includegraphics[width=1\linewidth]{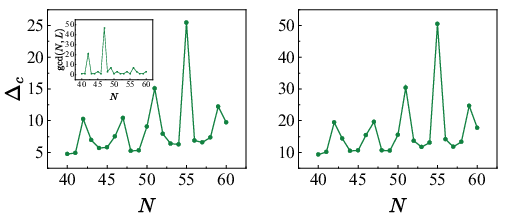}
    \caption{Dependence of the critical quasiperiodic potential strength $\Delta_c$ on the helical winding (long-range hopping range) $N$ for a chain of length $L=987$ at half filling, where $\Delta_c$ is extracted from the zero crossing of the geometric Binder cumulant. Left panel: $J_N=0.5$; right panel: $J_N=1.0$. Some pronounced spikes occur at $N$ values with $\gcd(N ,L)>1$the , inset graph corresponds to $\ gcd(N,L)$ vs $N$.
}
    \label{fig:4}
\end{figure}

\subsection{Dependence on helical range and commensurability effects}

We next fix the helical hopping amplitude and scan the winding (long-range hopping range) $N$.
Figure~\ref{fig:4} shows $\Delta_c$ as a function of $N$ for $N=40$--$60$ at half filling and $L=987$, for two values of the helical hopping amplitude: (left) $J_N=0.5$ and (right) $J_N=1.0$.
While $\Delta_c(N)$ exhibits an overall scale that increases with $J_N$ (as expected from Fig. \ref{fig:3}), its dependence on $N$ is strongly non-smooth: pronounced spikes appear at specific windings.
Notably, some of these spikes correlate with commensurate situations where $ \gcd(N,L)>1$, indicating that finite-size arithmetic relations between the winding and the ring length can enhance the effectiveness of the helical connectivity and shift the apparent transition. We will further investigate the $N$-dependence in the thermodynamic limit section. 

\begin{figure}
    \includegraphics[width=\linewidth]{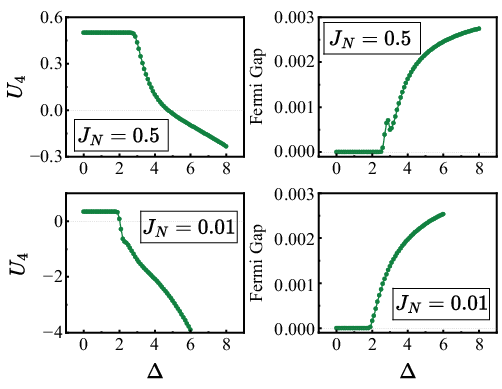}
    \caption{Comparison of the geometric Binder cumulant $U_4(\Delta)$ and the single-particle
   normalized Fermi gap
    $\tilde{\Delta}_{\rm sp}
    =(\epsilon_{N_p+1}-\epsilon_{N_p})/
    (\epsilon_{\max}-\epsilon_{\min})$, as functions of the quasiperiodic potential strength $\Delta$ for
    $L=987$. Top row:
    strong helical hopping $J_N=0.5$; bottom row: weak helical hopping $J_N=0.01$.
    In each row, $U_4$ is shown on the left and the Fermi gap on the right. The gap
    remains close to zero at small $\Delta$ and begins to open in the parameter
    region where $U_4$ departs from its extended-phase plateau.}
    \label{fig:5}
\end{figure}

Taken together, Figs. \ref{fig:3} and \ref{fig:4} separate two effects.
First, increasing $J_N$ produces a robust, monotonic stabilization of the extended phase.
Second, at fixed $J_N$ the winding $N$ can introduce strong commensurability-induced anomalies in finite systems, leading to sharp spikes in $\Delta_c(N)$.
In the following, we therefore repeat the extraction of $\Delta_c$ across multiple Fibonacci system sizes and perform finite-size scaling to determine which features persist in the thermodynamic limit and which are attributable to $(N,L)$ commensurability.

\subsection{Gap Analysis}
To complement the polarization-based diagnosis, we compare the geometric Binder
cumulant with a spectral indicator derived from the single-particle spectrum.
We consider the normalized single-particle Fermi gap
\[
\widetilde{\Delta}_{\rm sp}
\equiv
\frac{\epsilon_{N_p+1}-\epsilon_{N_p}}
{\epsilon_{\max}-\epsilon_{\min}},
\]
where the single-particle eigenvalues are ordered as
\(\epsilon_1 \leq \epsilon_2 \leq \cdots \leq \epsilon_L\).
Here \(\epsilon_{N_p}\) is the highest occupied single-particle level and
\(\epsilon_{N_p+1}\) is the lowest unoccupied level. The normalization by the
full single-particle bandwidth allows us to compare the gap on the same scale
for different values of \(\Delta\).

Fig. \ref{fig:5} displays \(U_4(\Delta)\) together with the normalized Fermi gap. We show
two representative helical hopping strengths: a strong helical coupling
\(J_N=0.5\) in the top row and a weak helical coupling \(J_N=0.01\) in the bottom
row. In each row, \(U_4\) is shown on the left and the normalized Fermi gap on
the right.

The Fermi gap remains close to zero for small quasiperiodic potential strength
\(\Delta\), consistent with the extended-phase regime where the spectrum near the
Fermi level is dense. As \(\Delta\) is increased, the gap begins to open in the
same broad parameter region where $U_4(\Delta)$ reshapes parts from its extended-phase
plateau. We emphasize that the gap onset does not necessarily coincide exactly
with the zero crossing of \(U_4\); rather, it provides a spectral consistency
check supporting the polarization-based localization diagnostic.

\begin{figure*}[t]
    \centering
    \includegraphics[width=0.5\textwidth]{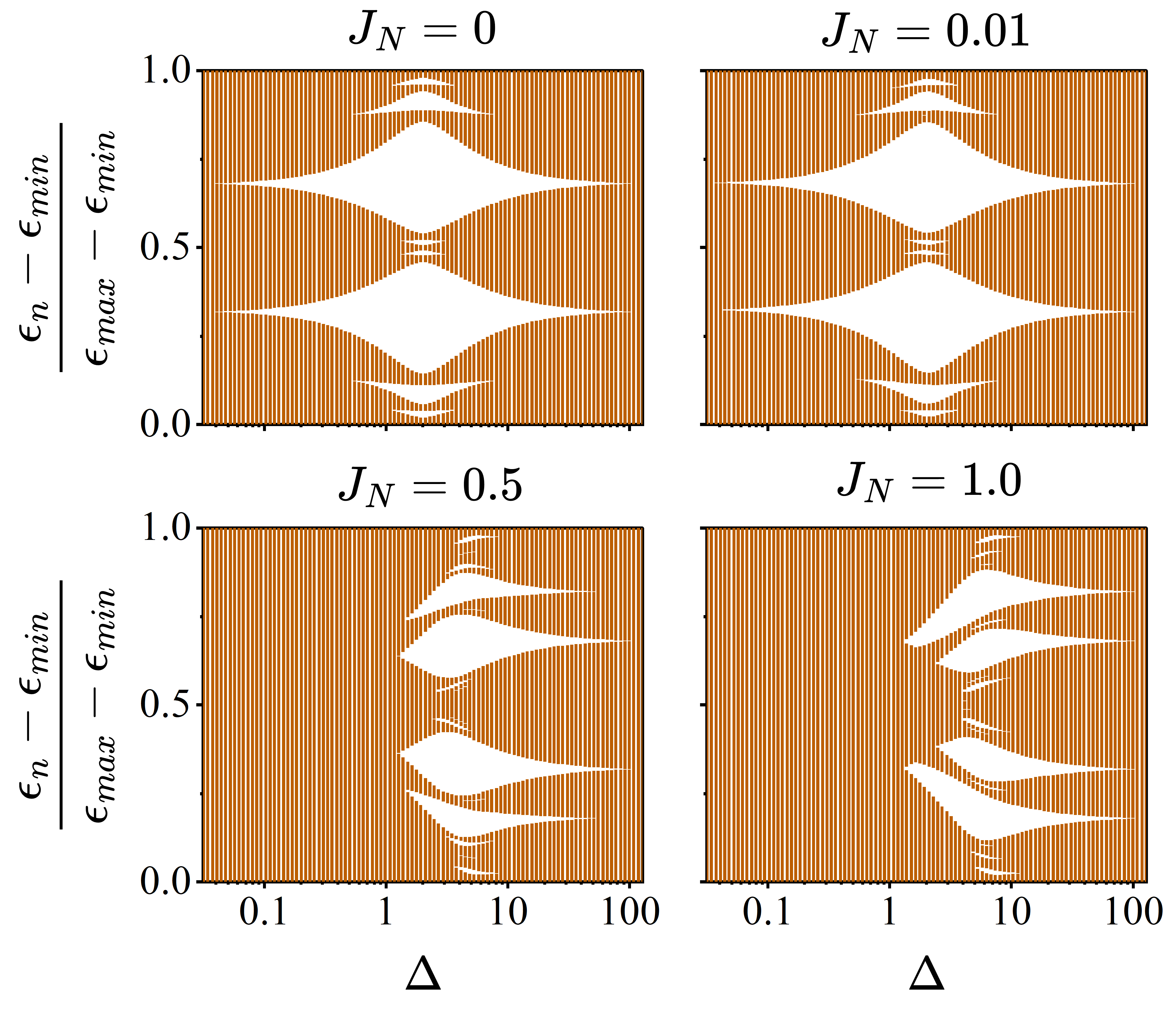}\hfill
    \includegraphics[width=0.5\textwidth]{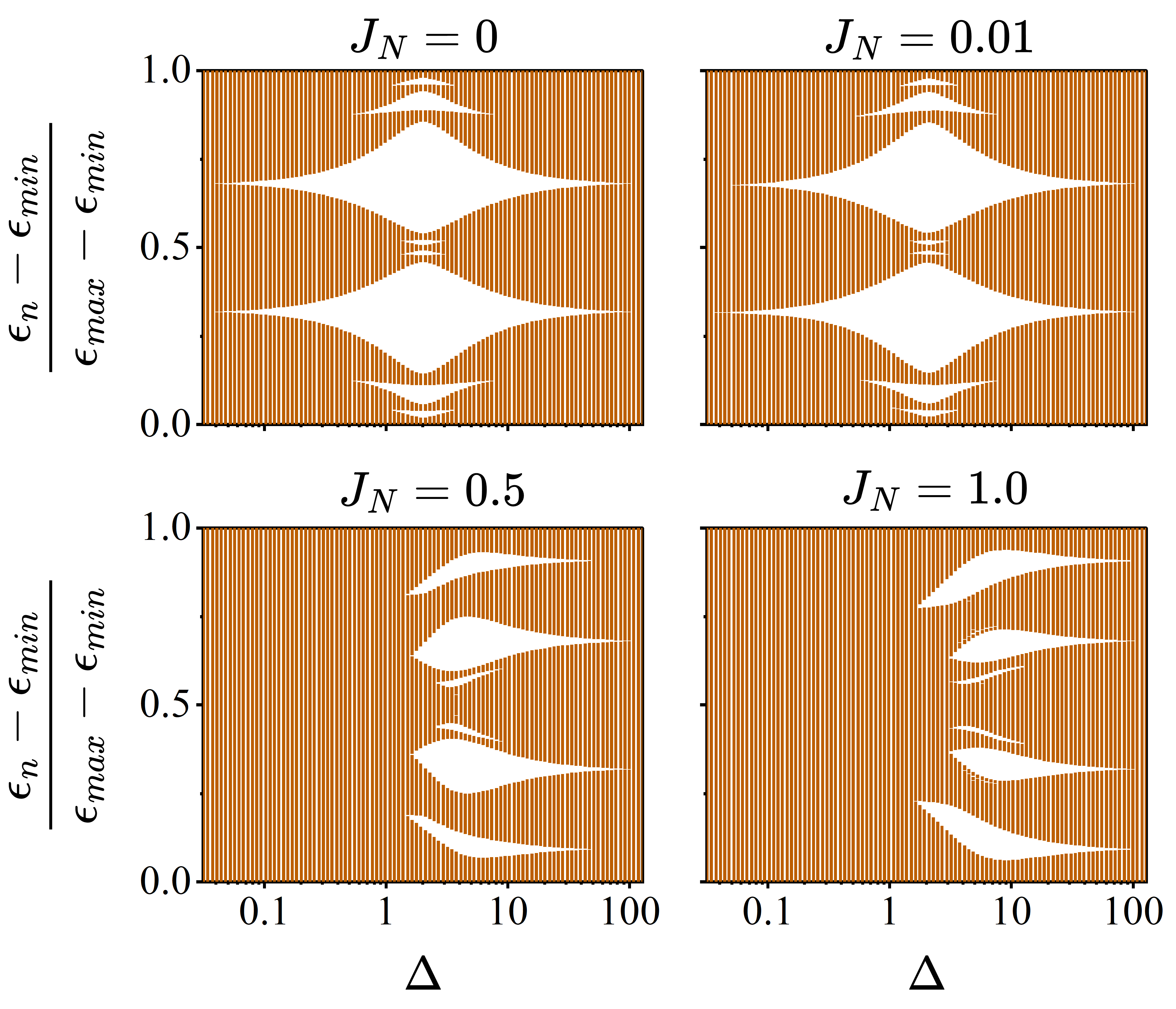}
    \caption{Normalized single-particle energy spectrum of the helical Aubry-Andr\'e model as a function of the quasiperiodic potential strength $\Delta$ (logarithmic scale on the horizontal axis) for $L=4181$.
    Each panel shows the rescaled eigenenergies
    $\tilde{\epsilon}_n=(\epsilon_n-\epsilon_{\min})/(\epsilon_{\max}-\epsilon_{\min})\in[0,1]$
    versus $\Delta$ for different helical hopping strengths ,$J_N$ (indicated above each, subplot).
    Left set: helical range $N=40$.
    Right set: helical range $N=100$.
    White regions correspond to parameter intervals with no eigenvalues (spectral gaps), illustrating how increasing $J_N$ reshape the gap structure.}
    \label{fig:6}
\end{figure*}

To place the Fermi-gap behavior in a broader context, we also examine how the entire single-particle spectrum changes with $\Delta$.
Fig. \ref{fig:6} plot the normalized energies
$\tilde{\epsilon}_n=(\epsilon_n-\epsilon_{\min})/(\epsilon_{\max}-\epsilon_{\min})\in[0,1]$
as a function of $\Delta$ (logarithmic horizontal axis) for $L=4181$.
White regions correspond to parameter intervals with no eigenvalues at the plotted normalized energy, i.e., spectral gaps in the rescaled spectrum.
Fig. \ref{fig:6} highlights the dependence on the helical hopping amplitude $J_N$ at fixed winding (left set $N=40$, right set $N=100$): increasing $J_N$ substantially reshapes the quasiperiodic band structure, shifting and splitting prominent bands and modifying the gap pattern across a wide range of $\Delta$.

Taken together, Figs. \ref{fig:5} and \ref{fig:6} show that the opening of the Fermi gap is broadly consistent with the localization crossover identified by $U_4(\Delta)$, while also revealing substantial spectral restructuring driven by helical tunneling.
Because Fermi gap is a local level-spacing measure that can fluctuate strongly at finite size, we use the geometric Binder cumulant as the primary tool for locating $\Delta_c$, and treat the gap analysis as an independent spectral check and as a guide to how $(J_N,N)$ reshape the quasiperiodic spectrum.

\section{Zeckendorf Decomposition and Thermodynamic Limit}

In quasiperiodic lattice models it is natural to study finite systems whose length is taken from the Fibonacci sequence,
$L=F_n$, because rational approximants of the incommensurate modulation are then compatible with periodic boundary
conditions. A important point is that, when the thermodynamic limit is taken along the Fibonacci subsequence,
specifying the filling only through a rational value $N_p/L$ (e.g. 1/2) does not uniquely determine how the
many-body sector is approached as $n\to\infty$. Different sequences of particle numbers $N(F_n)$ that all satisfy
$N_p(F_n)/F_n\to\nu$ can correspond to different sectors and may lead to distinct finite-size trends.

To define a controlled thermodynamic limit we  follow a Zeckendorf construction.
Every positive integer $N_p$ admits a unique Zeckendorf decomposition,
\begin{equation}
N_p=\sum_{\alpha} F_{\alpha}
\end{equation}
meaning that $N_p$ is written as a sum of nonconsecutive Fibonacci numbers. We choose a reference system size $L_0=F_{n_0}$
and a reference particle number $N_0$, compute its Zeckendorf indices $\{\alpha\}$, and then
generate particle numbers for larger Fibonacci sizes by an index shift:
\begin{equation}
L_m = F_{n_0+m},\qquad
N_m = \sum_{\alpha} F_{\alpha+m}
\end{equation}
This procedure keeps the digit pattern in the Zeckendorf representation fixed while increasing the system size, and
therefore defines a unique sequence of many-body sectors along the Fibonacci subsequence.

\begin{figure}
    \includegraphics[width=\linewidth]{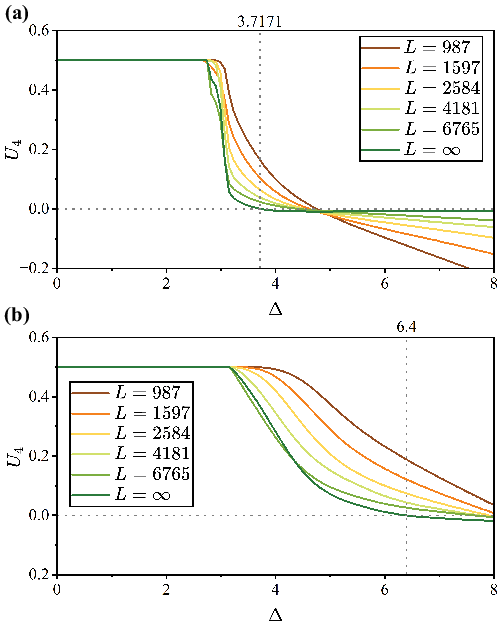}
    \caption{Geometric Binder cumulant $U_4$ as a function of quasiperiodic potential strength $\Delta$ for Fibonacci system sizes $L=987,1597,2584,4181,6765$ at the fixed-filling sequence defined by the Zeckendorf-shift construction where (a) $N=40$, $J_N=0.5$ (b) $N=100$, $J_N=0.5$. The dark-green curve shows the thermodynamic-limit estimate $L\to\infty$ obtained by extrapolating at fixed $\Delta$ (linear fit in $1/L$). The horizontal dashed line marks $U_4=0$, separating the delocalized regime ($U_4>0$) from the localized regime ($U_4<0$), and the vertical dashed line indicates the critical points.}
    \label{fig:7}
\end{figure}

The filling factor $\nu_m=N_m/L_m$ converges as $m\to\infty$ because $F_{n-k}/F_n\to \varphi^{-k}$, where
$\varphi=(1+\sqrt{5})/2$ is the golden ratio. More explicitly, writing $d_{\alpha}=n_0-\alpha$,
\begin{equation}
\nu_\infty = \lim_{m\to\infty}\frac{N_m}{F_{\alpha+m}}
= \sum_{\alpha} \varphi^{-d_\alpha},
\label{eq:nu_infty}
\end{equation}
which is generally irrational. Thus, Zeckendorf representation provides a well-defined thermodynamic limit
$F_{\alpha+m}\to\infty$ at fixed Zeckendorf filling ratio.

In the calculations below we fix the initial pair to $(L_0,N_0)=(987,494)$. The corresponding Zeckendorf decomposition is
\begin{equation}
494 = 377 + 89 + 21 + 5 + 2
= F_{14}+F_{11}+F_{8}+F_{5}+F_{3},
\end{equation}
and the shifted sequence yields the particle numbers
\begin{align}
(L,N)=&(987,494),(1597,799),(2584,1293),\\&(4181,2092),(6765,3385) \nonumber
\label{eq:LN_sequence}
\end{align}
which approach an asymptotic filling close to one half. Thermodynamic-limit
estimates are then obtained by comparing results across the Fibonacci sizes and
by extrapolating observables in powers of $1/L$.

\begin{figure}
    \centering
    \includegraphics[width=\linewidth]{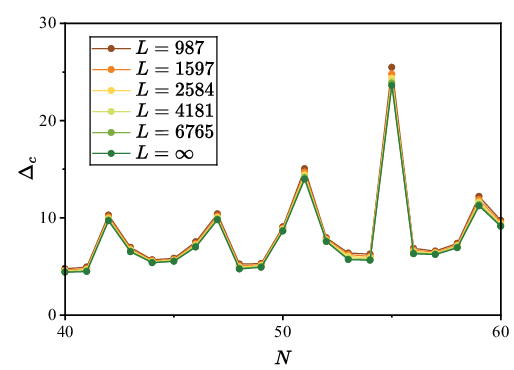}
    \caption{Critical quasiperiodic potential strength $\Delta_c$ as a function of the helical winding (long-range hopping range) $N$ for $N=40$--$60$ at fixed helical hopping amplitude $J_N=0.5$. The finite-size values are shown for Fibonacci system sizes $L=987$, $1597$, $2584$, $4181$, and $6765$, where $\Delta_c$ is extracted from the zero crossing of the geometric Binder cumulant $U_4(\Delta)$. The dark-green curve denotes the thermodynamic-limit estimate $L\to\infty$, obtained by extrapolating $\Delta_c$ linearly in $1/L$.}
    \label{fig:8}
\end{figure}

To illustrate how the Zeckendorf-shift sequence stabilizes the finite-size scaling, in Fig.~\ref{fig:7} we plot $U_4(\Delta)$ for the Fibonacci sizes $L=987,1597,2584,4181,$ and $6765$ using the shifted particle numbers in Eq.~(16), and construct an $L\to\infty$ estimate by extrapolating $U_4$ at fixed $\Delta$ with a linear fit in $1/L$ (dark-green curve). In both panels the finite-size curves approach the extrapolated thermodynamic-limit profile smoothly, showing that the Zeckendorf prescription yields a well-behaved sequence of many-body sectors for comparing different $L$. The horizontal dashed line marks $U_4=0$, separating the delocalized regime ($U_4>0$) from the localized regime ($U_4<0$), while the vertical dashed line indicates the corresponding thermodynamic-limit critical point extracted from the sign change of the extrapolated curve.

For the shorter helical range, $N=40$ [Fig.~\ref{fig:7}(a)], the thermodynamic-limit extrapolation gives $\Delta_c\simeq 3.72$. In the finite-size data one can identify an intermediate region in which $U_4$ decreases gradually before entering the clearly localized regime. However, this intermediate structure is progressively reduced with increasing $L$ and disappears in the $L\to\infty$ limit, where the extrapolated curve crosses into the localized regime without retaining a separate extended crossover window. By contrast, for the longer helical range, $N=100$ [Fig.~\ref{fig:7}(b)], the transition is shifted to a substantially larger value, $\Delta_c\simeq 6.4$, confirming that increasing the helical range further stabilizes the extended phase at fixed $J_N$. In this case the intermediate region does not vanish under extrapolation, but remains visible in the thermodynamic-limit curve, indicating that the longer-range helical connectivity sustains a broader crossover regime before full localization sets in.

To further clarify the role of the helical winding, we now examine the thermodynamic-limit behavior of $\Delta_c$ as a function of $N$ over the range $N=40$-$60$ at fixed $J_N=0.5$. Figure~8 shows the finite-size critical potentials extracted for the Fibonacci sizes $L=987,1597,2584,4181,$ and $6765$, together with the $L\to\infty$ estimate obtained from a linear extrapolation in $1/L$. In contrast to the stronger finite-size sensitivity suggested by the single-size scan, the five finite-size curves now lie very close to one another throughout the entire interval. This near-collapse indicates that once the many-body sectors are chosen consistently, the $N$-dependence of $\Delta_c$ is already well converged at the accessible Fibonacci sizes. Most importantly, the non-monotonic structure survives the extrapolation: the pronounced maximum at $N=55$ remains the dominant feature of the curve, and the smaller enhancements near $N=51$ and $N=59$ are also retained in the thermodynamic-limit estimate. We therefore conclude that the main spike structure is not an artifact of a particular finite size, but a robust property of the helical Aubry-Andr\'e model.

The physical origin of the spike structure can be understood from the way the helical hopping modifies the quasiperiodic potential. Since the long-range term connects sites $j$ and $j+N$, the onsite energy at the shifted site is
\[
v_{j+N}= \cos(2\pi \alpha j+2\pi \alpha N).
\]
Thus, varying the winding $N$ changes the phase shift $2\pi\alpha N$ with which the helical hop probes the Aubry-Andr\'e modulation. Special values of $N$ occur when this phase shift approaches resonant values. If $2\pi\alpha N \approx 0 \pmod{2\pi}$, then $v_{j+N}\approx v_j$, so the helical hop connects sites with nearly equal onsite energies. This enhances hybridization and makes the additional hopping channel particularly effective in stabilizing the extended phase, thereby shifting $\Delta_c$ to larger values. On the other hand, if $2\pi\alpha N \approx \pi \pmod{2\pi}$, then $\epsilon_{j+N}\approx -\epsilon_j$, so the hop connects sites with nearly opposite onsite energies. This corresponds to an anti-phase resonance, which can also strongly reorganize the low-energy states and produce an anomalously large critical potential. In this way, the spikes in $\Delta_c(N)$ can be interpreted as arising from resonant helical sampling of the quasiperiodic landscape.

\section{Conclusion}
We have studied localization in a helical extension of the Aubry-Andr\'e model with an additional $N$th-neighbor hopping term $J_N$. Using the geometric Binder cumulant $U_4$ under periodic boundary conditions, we located the delocalization--localization transition from the zero crossing of $U_4(\Delta)$. Our results show that increasing the helical hopping strength $J_N$ stabilizes the extended phase and shifts the critical quasiperiodic potential $\Delta_c$ to larger values. The comparison with the Fermi gap supports the same picture, since the gap opens in the same parameter regime where $U_4(\Delta)$ leaves its extended-phase behavior.

We also found that the dependence of $\Delta_c$ on the helical winding $N$ is strongly non-monotonic and displays pronounced spike-like anomalies. Finite-size scaling across Fibonacci system sizes shows that the dominant spike structure survives in the thermodynamic-limit extrapolation, indicating that these features are not merely artifacts of a particular finite size. The spikes can be understood as arising from resonant helical sampling of the quasiperiodic potential, since varying $N$ changes the phase shift with which the long-range hopping probes the Aubry-Andr\'e modulation.

Overall, the helical Aubry-Andr\'e model provides a simple setting in which quasiperiodicity, long-range hopping, and geometry compete in a nontrivial way. The geometric Binder cumulant proves to be an effective tool for tracking this competition and for determining the phase boundary under periodic boundary conditions.

\begin{acknowledgements}
This work is supported by the Scientific and Technological Research Council od T{\"u}rkiye (TUBITAK) under
Grant No. 125F435 and the Turkish Academy of Sciences
(TUBA) and Grant No. AD-2026.
    BH gratefully acknowledges support by HUN-REN 3410107 (HUN-REN-BME-BCE Quantum Technology Research Group), by the National Research, Development and Innovation Fund of Hungary within the Quantum Technology National Excellence Program (Project No. 2017-1.2.1-NKP-2017-00001), by Grants No. K142179, No. K142652, and No. FK142601 and by the BME-Nanotechnology FIKP Grant No. (BME FIKP-NAT).  
\end{acknowledgements}

\appendix
\section{DERIVATIONS}
\appendix
For discrete set of points we let $Z(k)=\langle e^{ikx}\rangle$ with $k_q=2\pi q/L$ and with $Z(0)=1$. Now we have 
\begin{align}
    M_2&=-\frac{d^2Z(k)}{dk^2}\approx-\frac{Z(k_1)+Z(-k_1)-2Z(0)}{k_1^2} \\
    &\approx \frac{-(Z_1+Z_{-1}-2Z_0)}{k_1^2}, \quad (Z_1+Z_{-1}=2|Z_1|)  \nonumber 
\end{align}
Hence we get 
\begin{align}
    M_2=\frac{-(2|Z_1|-2)}{k_1^2}&=\frac{-L^2}{4\pi^2}(2|z_1|-2) \\
    &=\frac{L^2}{2\pi^2}(1-|Z_1|) \nonumber
\end{align}
Now for $M_4$ we have 
\begin{align}
    \frac{d^4Z(k)}{dk^4}=\frac{Z(-2k_1)-4Z(-k_1)+6Z(0)-4Z(k_1)+Z(2k_1)}{k_1^4}
\end{align}
Then 
\begin{align}
    M_4&=\frac{Z_{-2}-4Z_{-1}+6Z_0-4Z_1+Z_2}{k_1^4} \\
    &=\frac{(Z_2+Z_2^*)-4(Z_1+Z_1^*)+6}{k_1^4} \nonumber
\end{align}
Therefore we get 
\begin{align}
    M_4&=\frac{2|Z_2|-8|Z_1|+6}{16\pi^4/L^4} \\
    &=\frac{L^4}{8\pi^4}(|Z_2|-4|Z_1|+3) \nonumber
\end{align}

\bibliography{bib}
\end{document}